\documentclass{aa}
\usepackage{graphicx}
\usepackage{longtable}
\usepackage{latexsym}
\usepackage{amssymb}
\usepackage{lscape}
\begin{document}
\title{
[CII] at 158 $\mu$m as a star formation tracer in late-type galaxies
}

\author{A. Boselli \inst{1}
\and G. Gavazzi \inst{2}
\and J. Lequeux \inst{3}
\and D. Pierini \inst{4}
}

\authorrunning{A.Boselli et al.}
\titlerunning{ The [CII] 158 $\mu$m line as a tracer of star formation in
late-type galaxies}

\offprints{Alessandro Boselli}

\institute{
Laboratoire d'Astrophysique de Marseille, BP 8,
Traverse du Siphon, F-13376 Marseille Cedex 12, France\\
\email {Alessandro.Boselli@astrsp-mrs.fr}
\and Universit\`a degli Studi di Milano-Bicocca, Dipartimento di Fisica,
Piazza dell'Ateneo Nuovo 1, 20126 Milano, Italy\\
\email {Giuseppe.Gavazzi@mib.infn.it}
\and DEMIRM and URA 336 du CNRS, Observatoire de Paris, 61 Av. de
l'Observatoire, F-75014 Paris, France \\
\email {James.Lequeux@obspm.fr}
\and Ritter Astrophysical Research Center, The University of Toledo, 2801
West Bancroft, Toledo, OH 43606, USA\\
\email {pierini@ancona.astro.utoledo.edu}
}

\date{}

\abstract{We present a calibration of the massive star formation rate
  vs. [CII] luminosity relation based on a sample of nearby, late-type
  galaxies observed with ISO-LWS and imaged in the H$\alpha$+[NII] line.
  The relation holds for far-IR luminosities
  $10^8 \leq L_{FIR} \le 10^{10.5} L _\odot$.
  The derived star formation rates have an uncertainty of about a factor of 10.
  Part of this uncertainty is due to the different mix of contributions
  to the [CII] emission from the different components of the interstellar
  medium in individual galaxies, as discussed in an appendix.
}

\maketitle
%

\section{Introduction}

Galaxies have transformed their primordial gas, mostly
atomic hydrogen, into stars at a rate which can be variable with time.
During merging with nearby companions, which were probably frequent
in the high density environments of the early Universe,
most of the gas might have been efficiently transformed into stars on short
timescales ($\leq$ 10$^8$ yr), forming elliptical galaxies.
In spiral galaxies in the local Universe,
the activity of star formation seems to be governed by the gas surface
density, possibly modulated by galactic differential rotation (Kennicutt
1998a). The resulting star formation history of
these objects should have been thus monotonic,
governed by some of the physical
properties of the original protogalaxy such as its angular momentum
(Sandage 1986) or total mass (Boselli et al. 2001).  \\
In order to understand the formation and evolution of galaxies, it is thus
necessary to study the activity of
star formation in objects of different types at different epochs.
Quantifing the rate of star formation of normal, late-type galaxies in the
local Universe is quite straightforward using H$\alpha$ and UV data
together with stellar population synthesis models (Kennicutt 1998a,b),
once a correction for dust attenuation is made.
This is not so easy for galaxies at higher redshift.
Deep optical observations will provide UV rest-frame photometry, but
the H$\alpha$ line is shifted outside the optical domain for $z$$\geq$0.5,
while remaining in principle accessible in the near-IR up to $z$ $\sim$ 6.
However the wide-field spectroscopic
facilities suitable for statistical analysis which will be available
in the near future, such as VIRMOS on the VLT, will be limited to $\lambda$
$\leq$ 1.8 $\mu$m, thus to $z$ $\sim$ 1.7. \\
Another problem is that a large fraction of galaxies are expected to form
stars in violent starbursts at $z$ $\geq$ 1 (Steidel et al. 1999). The
starburst regions are characterized by a high extinction, so that UV and
Balmer lines cannot be reliably used: other indicators such as the
far-IR luminosity should then be used (Kennicutt 1998a, b).\\
Given the low spatial resolution of available far-IR/submm data ($\sim$ 1
arcmin for
IRAS and ISO at 100 $\mu$m,
$\sim$ 15 arcsec for SCUBA), confusion is important at high redshifts.
Mid-IR (MIR) surveys such as those performed with ISOCAM provide data
with a higher spatial resolution ($\sim$ 6 arcsec),
but the relation between MIR luminosity and star formation is still
uncertain (Boselli et al. 1997a; 1998; Roussel et al. 2001).\\
Some far-IR forbidden lines, in particular the [CII] 158 $\mu$m one, are
potentially better indicators of star formation in distant galaxies (Stacey
et al. 1991a).
The [CII] line is one of the main coolants of the
interstellar gas. Heating of the interstellar
  gas is mainly due to photo--electrons emitted by dust grains and
  Polycyclic Aromatic Hydrocarbons (PAHs) submitted to ultraviolet radiation
  from stars (Bakes \& Tielens 1994), both in the diffuse interstellar medium
  (ISM) (Wolfire et al. 1995) and in Photodissociation Regions (PDRs), at the
  interfaces between molecular clouds and HII regions (Tielens \& Hollenbach
  1985; Bakes \& Tielens 1998). The photoelectric effect is due essentially to
  photons with $\rm 6 \le h \nu < 13.6~eV$. In the field of a galaxy, this
  radiation is dominated by B3 to B0 stars with $\rm 5 \leq M \leq 20~M_{\sun}$
  (e.g. Xu et al. 1994) but of course hotter, more massive stars can also
  contribute locally.
However we cannot expect the intensity of the [CII] line from a
galaxy to be simply proportional to the star formation rate (SFR) and to
the intensity of the H$\alpha$ line. This is due not only to the variety of
[CII] line sources with their different physical conditions,
which will be discussed
further in the appendix, but also to the fact that the excitation of the upper
fine-structure level of [CII] saturates at high temperatures and at high
densities, as discussed e.g. by Kaufman et al. (1999; see their Fig. 2 and
3). Fortunately the [CII] line is only affected by extinction in extreme
starbursts (Luhman et al. 1998; Helou
2000) or in edge-on galaxies (Heiles 1994), being not very optically thick
($\tau \approx$ 1 in the Orion PDR: Stacey et al. 1991b, Boreiko \& Betz
1996).\\
To summarize, the [CII] line cannot be a quantitatively accurate indicator
of star formation in any galactic or extragalactic environment.
However it is very strong, and next generation far-IR and
submillimeter facilities such as FIRST-Herschel or ALMA will provide [CII]
line intensities for large samples of galaxies at different redshifts
(Stark 1997). In particular, the high angular resolution of ALMA will
overcome the problem of confusion at high redshifts. It is thus useful to
examine empirically to which extent the [CII] line can be used as a star
formation indicator, and to calibrate it in terms of SFR in the local
Universe using other tracers.\\
This is the aim of the present paper, which does not attempt to discuss in
detail the physics of the emission in the [CII] line. We compare ISO-LWS
[CII] to H$\alpha$ data for a sample of 26 nearby galaxies (Sect. 2).
Available complementary data are then used to empirically calibrate
a relation between the [CII] luminosity and the SFR (Sect. 3).
We emphasize that this calibration is limited to relatively quiescent galaxies.
Sect. 4 contains a short dicussion and the conclusions. The relations
between the far-IR emission of different components of the interstellar
matter and their [CII] line emission are summarized in the Appendix.

\section {The sample}

An empirical calibration of the [CII] line as a star formation tracer
can be obtained by comparing H$\alpha$ and [CII] luminosities measured in
the same aperture. For this reason
  we consider a sample of 18 Virgo galaxies
observed with ISO-LWS.

The [CII] data are taken from Leech et al. (1999) and Smith \& Madden (1997),
  while H$\alpha$+[NII] images are available from Boselli \& Gavazzi
  (2002), Gavazzi et al. (2002) and Boselli et al. (2002). To increase the
statistics, we add M82,
  which was imaged in the H$\alpha$+[NII] line by Boselli \& Gavazzi
  (2002), NGC1569, whose $\rm H \alpha$ data are kindly made available to us
  by D. Bomans, and 6 objects (VCC 89, VCC 491 and VCC 1516, NGC 7469 and
7714, IC 4662) whose
  $\rm H \alpha$ data come from aperture photometry but
whose optical disc, given their small angular size, has been covered
to more than 40 \% by the [CII] ISO-LWS beam.\\
The final sample, which comprises the 26 galaxies listed in Table 1, is
not complete in any sense.\\
Far-IR surveys at high redshifts will preferentially
detect FIR-luminous galaxies, whose physical properties might
significantly differ from those of the normal, late-type nearby galaxies
analysed here. In our sample only two objects are luminous in the far-IR,
M82 ($\log L_{FIR}$ $=$ 10.51 L$_\odot$) and NGC 7469 ($\log L_{FIR}$ $=$
11.37 L$_\odot$),
the latter being an active galaxy which should be considered
with caution. We decided to add these two objects to test whether the present
results can be extended to far-IR luminosities higher than
$L_{FIR}$ $\ge$ 10$^{10.5}$ L$_\odot$.\\
The far-IR quiescent galaxies analysed here have
10$^8$ $\leq$ $L_{FIR}$ $\le$ 10$^{10.5}$ L$_\odot$.
Unless specified, all the results presented in this work are only valid for
galaxies in this range of far-IR luminosity.

\subsection {The data}

The available data for the galaxies analyzed here are listed in Table 1,
arranged as follow:

\begin {itemize}
\item {Column 1: VCC denomination for the Virgo cluster galaxies (Binggeli
et al.
1985), Messier,
NGC or IC name for the other objects.}
\item {Column 2: morphological type, from the VCC for Virgo galaxies, from NED
for the other objects.}
\item {Column 3: distance [Mpc], from Gavazzi et al. (1999) for Virgo,
from the catalogue of Tully (1988) for the other galaxies, or from the
redshift assuming $H_o$ = 75 $\rm{km ~s^{-1} Mpc^{-1}}$.}
\item {Column 4: absolute B magnitude, as determined from our own
photometry for the
Virgo galaxies, from NED for the other objects.}
\item {Column 5: H band luminosity, in solar units, defined as
$\log L_H = 11.36 -0.4 \times H_T +2\times \log(D)$, where $H_T$ is the total
H band magnitude and $D$ the distance in Mpc. For the Virgo galaxies, total
extrapolated near-IR magnitudes have been determined as described in Gavazzi
et al. (2000) using the compilation of Boselli et al. (1997b; 2000) and Gavazzi
et al. (1996a). K$'$ band magnitudes have been transformed into H magnitudes
adopting
a constant $H-K'$=0.25 (independent of type; see Gavazzi et al. 2000)
when the colour index is not available. For the other galaxies H magnitudes
are from
2MASS or from aperture photometry assuming an H band growth curve similar
to the
B band one.}
\item {Column 6, 7 and 8: (H$\alpha$ + [NII])$E.W.$ [\AA], flux
(- log, in $\rm{erg ~cm^{-2} s^{-1}}$) extracted in the same circular
apertures as LWS (i.e. 80 arcsec),
and reference; the fluxes and equivalent widths from  Kennicutt \& Kent (1983)
have been increased by 16 \%
as specified in Kennicutt et al. (1994) (also Kennicutt, private
communication);
when the equivalent widths are not given, we estimate the underlying
continuum from R band photometry using the relation: $R$ = 0.0 mag =
1.74 $\times$ 10$^{-9}$ erg cm$^{-2}$s$^{-1}$\AA$^{-1}$.}
\item {Column 9, 10 and 11: [CII] line flux,
(- log, in $\rm{erg ~cm^{-2} s^{-1}}$), equivalent width ([$\mu$m]) and
reference. Fluxes and
equivalent widths are all in the 80 arcsec circular apertures of LWS aboard
ISO.}
\item {Column 12: logarithm of the FIR luminosity, in solar units (L$_\odot$
= 3.83
$\times$ 10$^{26}$ W), $L_{FIR}$ = 4$\pi$$D^2$$\times$1.4$\times$$S_{FIR}$,
where
$S_{FIR}$ is the far-IR flux in W m$^{-2}$ between 42.5 and 122.5 $\mu$m,
$S_{FIR}$ = 1.26 $\times$ 10$^{-14}$ $\times$ (2.58 $\times$ $f_{60}$ $+$
$f_{100}$),
where $f_{60}$ and $f_{100}$ are the IRAS fluxes at 60 and 100 $\mu$m
respectively [Jy] (Helou et al. 1985).
Far-IR data have been taken from different
compilations for the Virgo galaxies (Thuan \& Sauvage 1992; Soifer et al.
1989, Rush et al. 1993,
and references therein), and from NED for the other galaxies.}
\item {Column 13: comments on individual objects, from NED: 1= starburst,
2= LINER, 3= AGN.}
\end{itemize}

\begin{table*}
\caption{The sample galaxies}
\label{Tab1}
\[
\begin{array}{rccccccccccrc}
\hline
\noalign{\smallskip}
{\rm Name}  & {\rm type} & {\rm dist} & M_B & L_H & ({\rm H}\alpha+{\rm
[NII]})E.W. & f({\rm H}\alpha+{\rm [NII]}) & {\rm ref.}
& F({\rm [CII]}) & {\rm [CII]}E.W. & {\rm ref.} & L_{FIR} & {\rm comm.} \\
      &      & \rm{Mpc} & \rm{mag} & \rm{L_\odot}& \rm{\AA} & \rm{erg~
cm^{-2}~ s^{-1}}& &\rm{erg~
cm^{-2}~ s^{-1}}
&\rm{\mu m} & &  \rm{L_\odot}& \\
\noalign{\smallskip}
(1) & (2) & (3) & (4) & (5) & (6) & (7) & (8) & (9) & (10) & (11) & (12) & (13) \\
\hline
\noalign{\smallskip}
VCC    66   & Sc&   17& -19.93& 10.22 & 24  &12.20  &  1&  12.11& 1.16 &1&
9.31& \\
VCC    89   & Sc&   32& -20.05& 10.65 & 23^a&11.94^a&  1&  12.04& 0.57 &2&
9.99& \\
VCC    92   & Sb&   17& -21.28& 11.03 &  5  &12.33  &  2&  12.06& 0.51 &1&
9.82&3\\
VCC   460   & Sa&   17& -19.98& 10.84 &  9  &12.33  &  6&  12.57& 0.23 &1&
9.52&2\\
VCC   465   & Sc&   17& -19.46&  9.88 & 73  &11.71  &  3&  12.08& 0.76 &2&
9.29& \\
VCC   491   &Scd&   17& -18.00&  9.42 & 74  &11.72  &  4&  12.17& 0.85 &2&
9.26& \\
VCC   857   & Sb&   17& -19.24& 10.53 &  1  &12.64  &  2&  12.85& 0.22 &1&
8.99&2\\
VCC   873   & Sc&   17& -19.41& 10.39 & 11  &12.44  &  5&  11.76& 0.87 &1&
9.68& \\
VCC  1003   &S0a&   17& -20.57& 11.04 &  5  &12.11  &  2&  12.85& 0.28 &1&
9.08&2\\
VCC  1043   & Sb&   17& -20.59& 10.85 &  7  &11.88  &  6&  12.09& 0.73 &1&
9.50&2\\
VCC  1110   &Sab&   17& -20.46& 10.96 &  2  &12.56  &  6&  12.68& 0.30 &1&
9.28&2\\
VCC  1158   & Sa&   17& -19.45& 10.54 &  0  &  -    &  2& <12.70&<0.87 &1&
<7.97& \\
VCC  1326   & Sa&   17& -17.96&  9.78 &  0  &  -    &  2&  12.89& 0.33 &1&
9.20& \\
VCC  1412   & Sa&   17& -19.42& 10.54 &  0  &  -    &  2& <12.89&<0.50 &1&
<8.07& \\
VCC  1516   &Sbc&   17& -19.01&  9.85 & 10^a&12.29^a&  6&  12.09& 0.81 &2&
9.06& \\
VCC  1690   &Sab&   17& -21.47& 11.07 &  3  &12.31  &  2&  11.82& 0.59 &1&
9.90&3\\
VCC  1727   &Sab&   17& -20.81& 11.11 &  4  &12.08  &  1&  12.11& 0.63 &1&
9.73&2\\
VCC  1813   & Sa&   17& -19.77& 10.84 &  0  &  -    &  2& <12.70&<0.65 &1&
8.58& \\
VCC  1972   & Sc&   17& -19.27& 10.50 & 14  &11.79  &  4&  11.72& 0.79 &1&
9.65& \\
VCC  1987   & Sc&   17& -20.32& 10.66 & 24  &11.73  &  4&  11.55& 0.96 &1&
10.04& \\
VCC  2070   & Sa&   17& -20.01& 10.76 &  2  &12.34  &  4&  12.89& 0.29 &1&
8.72&3\\
\hline
   M82   &Pec& 3.63& -18.94& 10.69 & 70  &10.35  &  2& 9.86& 1.15 &4& 10.51&1\\
NGC1569  &IBm& 1.6 & -16.60&  8.52 &185  &10.84  &  8&11.22& 3.00 &3&  8.34&1\\
NGC7469  & Sa& 65.2& -21.43& 11.18 & 97^a&11.50^a&  7&11.64& 0.86 &5& 11.37&3\\
NGC7714  &Pec& 36.9& -20.24& 10.76 &150^a&11.50^a&  7&11.70& 2.14 &5& 10.44&2\\
IC4662   & Im&	  3& -16.10&  8.78 & 59^a&11.04^a&  1&11.90& 1.04 &3&  8.18& \\
\noalign{\smallskip}
\hline
\end{array}
\]
a: total values (galaxies whose optical sizes are comparable with the
ISO-LWS beam).

References of H$\alpha$+[NII] data:
1: Kennicutt\& Kent (1983);
2: Boselli \& Gavazzi (2002);
3: Gavazzi et al. (2002);
4: Romanishin (1990);
5: Boselli et al. (2002);
6: Young et al. (1996);
7: Kennicutt et al. (1987);
8: Bomans, private communication;
for references 1, 4 and 5 the fluxes and E.W.s are calculated in a 80
arcsec diameter
aperture form the H$\alpha$+[NII] images from
Boselli \& Gavazzi (2002), Gavazzi et al. (2002) and Boselli et al. (2002).

References of the [CII] line data:
1: Leech et al. (1999);
2: Smith \& Madden (1996);
3: Hunter et al. (2001);
4: Colbert et al. (1999);
5 Negishi et al. (2001), combined with Onaka, private communication.
\end{table*}

\noindent
The H band luminosity is a good tracer of the total dynamical mass of a galaxy
(Gavazzi et al. 1996b): the mass to near-IR light ratio was found to be
independent of mass for galaxies
spanning the whole range in morphological type from Sa to Im and BCD.
By normalizing the [CII] luminosity to the near-IR luminosity (within the same
circular aperture)
we remove
the well known luminosity-luminosity or luminosity-mass scaling
relations in a better way than by using optical luminosities.\\
However, since the aim of the present work is to establish a star formation
tracer directely comparable to those which will be available for
high-redshift galaxies
from far-IR/submillimeter  facilities, we will also use as a normalizer
the continuum underlying the [CII] line. This leads to the determination of
[CII] equivalent widths
([CII]$E.W.$, in $\mu$m). An advantage of using the continuum as a
normalizer is that it is obtained in the same aperture of the [CII] line.
A disadvantage is that this continuum is poorly determined, at least in the
ISO-LWS
observations (see Hunter et al. 2001).

\noindent
H$\alpha$ luminosities can be estimated from H$\alpha$+[NII] fluxes after
correcting for contamination by the [NII] lines and for extinction. Given
the good correlation observed between the [NII]/H$\alpha$ ratios and $L_H$
observed by Gavazzi et al. (in prep.), we use the statistical correction:
log ([NII]/H$\alpha$) = 0.35 $\times \log L_H - 3.85$.\\
We use the same extinction correction as proposed by Boselli et al. (2001),
namely
$A({\rm H}\alpha)$=1.1 magnitudes for Sa$\leq$Scd and Pec galaxies, and
$A({\rm H}\alpha)$=0.6 mag for the other objects,
with the exception of M82 for which we take $A({\rm H}\alpha)$=3 mag
(Gavazzi et
al., 2002,
in preparation).

\begin{figure}[!h]
\centerline {}
\vbox{\null\vskip 20.0 cm
\includegraphics{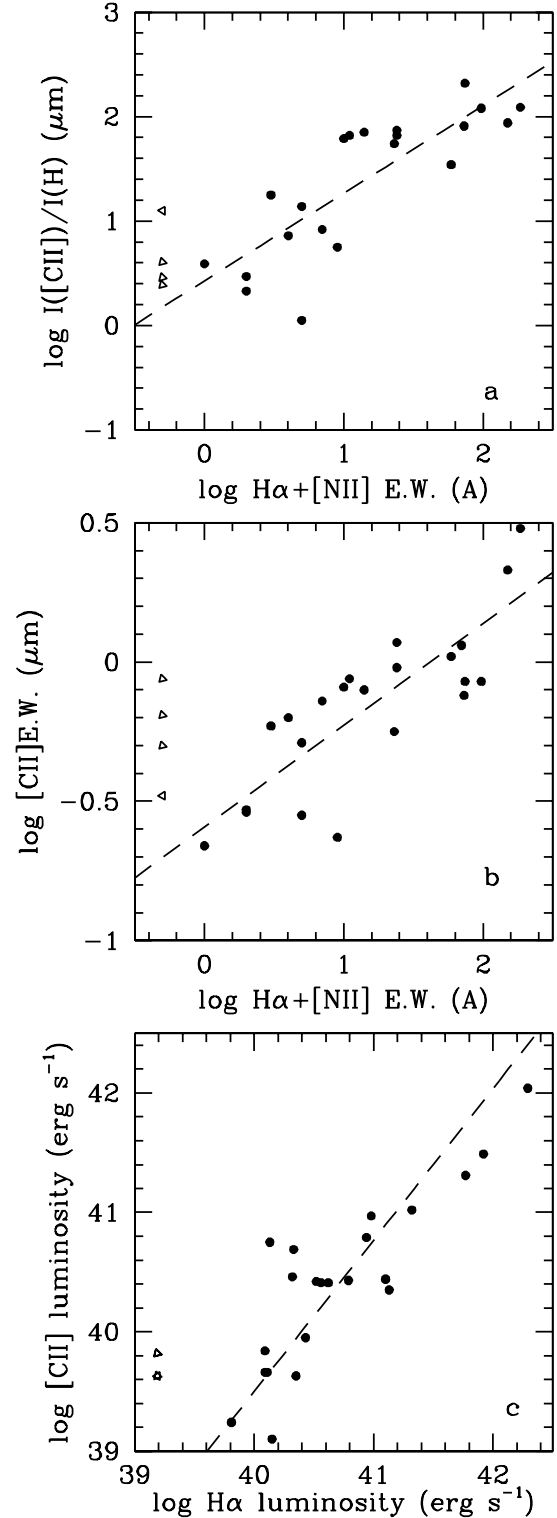}
}
\caption{The relationship between a) the [CII] line flux normalized to the
H band flux and the (H$\alpha$ + [NII])$E.W.$,
b) the [CII] and the (H$\alpha$ + [NII])$E.W.$ and c) the [CII] and
H$\alpha$ luminosities.
The dashed lines give the best fit to the data (eq. 2 and Table 2)
Open triangles are for [CII] and/or H$\alpha$
undetected galaxies. We adopt arbitrary upper limits for the
(H$\alpha$ + [NII])$E.W.$
and for the H$\alpha$ luminosity of 0.5 \AA ~and 1.6 10$^{39}$ erg s$^{-1}$
respectively.}
\label{fig.1}
\end{figure}

\section {Analysis}

In the following we use the formalism of Boselli et al. (2001).\\
The H$\alpha$ luminosity gives a direct measure of the global
photoionization rate of the
interstellar medium due to high mass ($m$ $>$ 10 ${\rm M_\odot}$), young
($\le$ 10$^7$ years) O-B stars (Kennicutt 1983; 1998a;
Kennicutt et al. 1994).
The total SFR can be determined by extrapolating the high-mass SFR to
lower mass stars using an initial mass function (IMF) with a given
slope $\alpha$ and upper and lower mass cutoffs $M_{up}$ and $M_{low}$
respectively.
Assuming that $L_{{\rm H}\alpha}$ is proportional to the SFR, then we have:

\begin{equation}
{SFR_{{\rm H}\alpha} ({\rm M_\odot yr^{-1}}) =
K_{{\rm H}\alpha}(\alpha,M_{up},M_{low}) \times L_{{\rm H}\alpha}}
\end{equation}

\noindent
where $K_{{\rm H}\alpha}(\alpha,M_{up},M_{low})$ is a proportionality
constant between the
H$\alpha$ luminosity $L_{{\rm H}\alpha}$ (in erg s$^{-1}$) and the
$SFR_{{\rm H}\alpha}$
(in M$_\odot$ yr$^{-1}$). The value of
$K_{{\rm H}\alpha}(\alpha,M_{up},M_{low})$,
as a function of the slope $\alpha$ and of the upper and lower mass cutoffs
$M_{up}$ and $M_{low}$ of the IMF,
can be determined from stellar population synthesis models.\\
We plot the relationship between the [CII] line flux normalized
to the H band (Fig. 1a) or the [CII]$E.W.$ (Fig. 1b), and the (H$\alpha$ +
[NII])$E.W.$.
Normalized entities are used here to remove the first order dependence on
luminosity; H magnitudes used in the normalization and H$\alpha$ +
[NII] parameters are determined within apertures comparable to the
beam of the [CII] observations. H magnitudes ($Hmag$) are transformed into
fluxes using the relationship:
$I(H)$ = 1.03 $\times$ 10$^6$$\times$10$^{-Hmag/2.5}$ (mJy)
(the resulting ratio $I([CII])$/$I(H)$ plotted in Fig. 1 is in units of
$\mu$m).

\noindent
Both figures show a strong correlation between the normalized [CII] line
and the (H$\alpha$ + [NII])$E.W.$ The relationship is shared by quiescent
galaxies
as well as by far-IR luminous objects, and is valid in the range
0 $\leq$ (H$\alpha$ + [NII])$E.W.$ $\leq$ 200 \AA, which encompasses the
whole normal,
late-type galaxies population in the local Universe. This result
  confrims the expectations of Pierini et al. (1999).\\
Figure 1c shows the relationship between the luminosities in the [CII]
and H$\alpha$ lines, the latter being corrected
for extinction and [NII] contamination as described in Sect. 2.1.

\noindent
This correlation is dominated by the scaling effect (bigger
galaxies have more of everything). The correlation observed for the
normalized
entities shown in Fig. 1a and b, however, confirm that the excitation
of the [CII] line is physically associated with the process of star
formation, or
more directly with the number of ionizing photons. 
This justifies the use of
the [CII] line as a tracer of star formation.The best fit to the data
for normalized entities are given in Table 2.\\
Figure 1c can be used to obtain the SFR from the [CII] luminosity.
The best fit to the data gives:

\begin{equation}
{\log L_{{\rm H}\alpha} = 8.875 (\pm 0.326) + 0.788 (\pm 0.098) \times \log
L_{[CII]}}
\end{equation}

\begin{table*}
\caption{Best fit to the relations between [CII] normalized entities
and the H$\alpha$+[NII] E.W. (\AA) (in logarithmic scales).}
\label{Tab2}
\[
\begin{array}{lcrcc}
\hline
\noalign{\smallskip}
{\rm Variable}& {\rm slope} & {\rm constant} & {\rm n.~of~objects} & {\rm R^{2a}}  \\
\noalign{\smallskip}
\hline
\noalign{\smallskip}
[CII]E.W. (\mu m) & 0.365 \pm 0.056& -0.592 \pm 0.169& 22& 0.68 \\
I([CII])/I(H) (\mu m) & 0.840 \pm 0.134& 0.427 \pm 0.310 & 21& 0.67 \\
\noalign{\smallskip}
\hline
\end{array}
\]
a: regression coefficient
\end{table*}

\noindent
as determined on the 22 detected objects with a $R^2$=0.76.\\


\noindent
The dispersion in the relation is high at low luminosities. We also remark
that 20 out of the 22 galaxies used to estimate this relation are quiescent
in the far-IR, i.e. $\log L_{FIR}$ $\leq$ 10.50 L$_\odot$.\\
Inserting $K_{{\rm H}\alpha}(\alpha,M_{up},M_{low})$ for different IMFs in
eq. (1),
[CII] luminosities can be used to derive SFRs in M$_\odot$ yr$^{-1}$ using
eq. (2).
For a Salpeter IMF ($\alpha$=2.35) in the mass range between 0.1 and 100
M$_\odot$, the population
synthesis models of Kennicutt (1998b) give $K_{{\rm
H}\alpha}(\alpha,M_{up},M_{low})$ = 1/1.26 10$^{41}$
({\rm M$_\odot$ yr$^{-1}$/erg s$^{-1}$}), thus:

\begin{equation}
{SFR = 5.952~ 10^{-33} \times 10^{0.788 \times \log L_{[CII]}} ~~~~~~~ {\rm
M}_\odot {\rm yr}^{-1}}
\end{equation}



\noindent
from eq.(2), where $L_{\rm [CII]}$ is in erg s$^{-1}$.
Values of $K_{{\rm H}\alpha}(\alpha,M_{up},M_{low})$ for different
IMFs can be found in Boselli et al. (2001).\\
In order to characterize the properties of the selected sample, thus to
understand
how universal this calibration is, we study the relation between the [CII]
line
emission and two parameters representative of the general properties of the
target
galaxies, the H band luminosity (as a tracer of the total mass)
and the morphological type (Fig. 2).

\begin{figure*}
\centerline {}
\vbox{\null\vskip 15. cm
\includegraphics{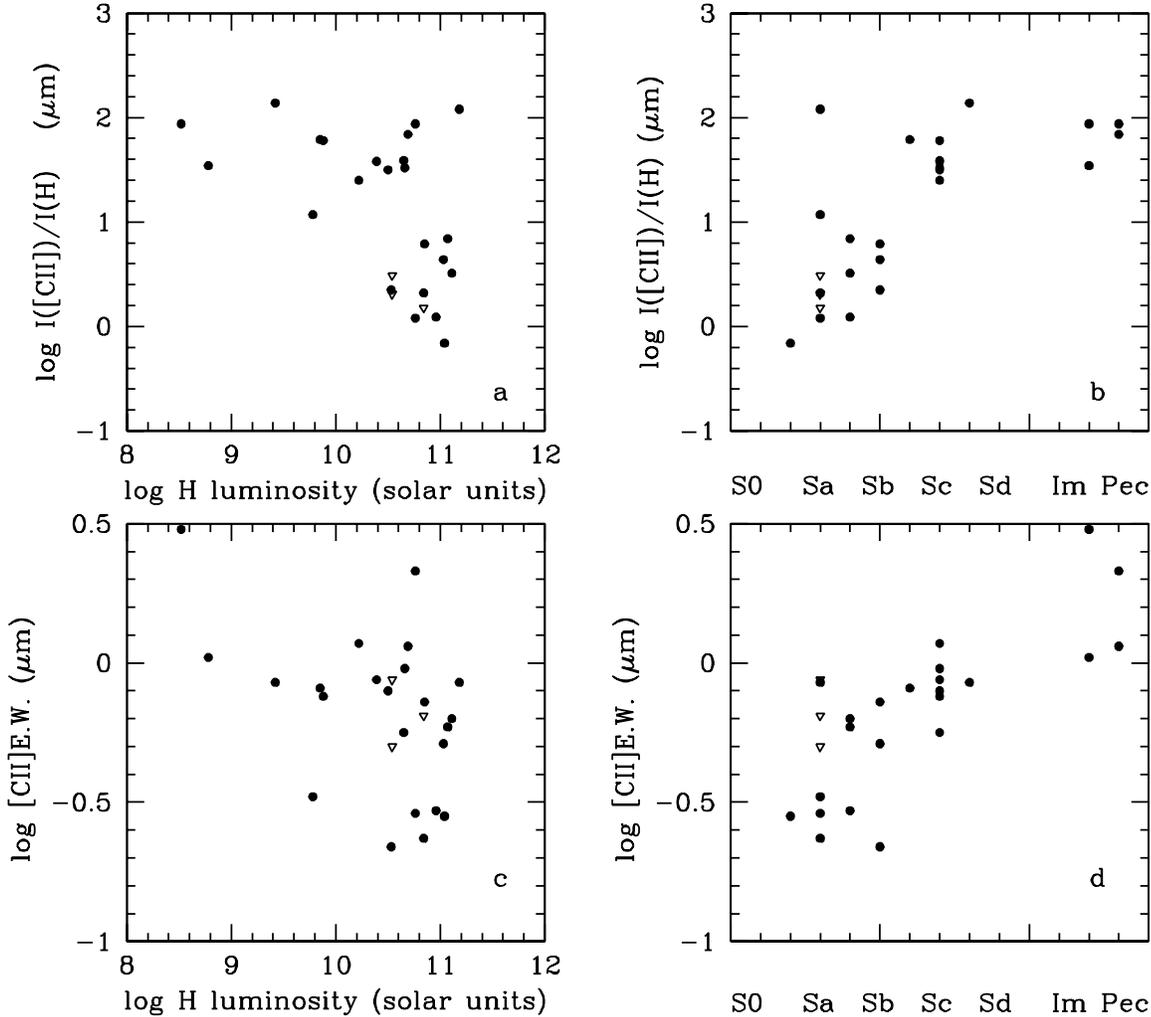}}
\caption{The relationship between the [CII] line flux normalized to the
H band flux and a) the H band luminosity, and b) the morphological type. The
relationship between
the [CII] $E.W.$ and c) the H band luminosity, and d) the morphological type.
Symbols as in Fig. 1.
}
\label{fig.2}
\end{figure*}

\noindent
Figure 2 shows that massive, early type spirals have in general a lower
normalized [CII]
emission than late-type, low mass ones, in agreement with
Leech et al. (1999). This behaviour is identical to
that observed for the star formation rate in the nearby Universe,
as discussed in Boselli et al. (2001):
massive galaxies form less stars per unit mass than low-mass objects.\\
We do not observe any particular relationship between two pure
far-IR indices which will be easily observable in deep,
far-IR surveys of the Universe, the
[CII] $E.W.$ and the FIR luminosity, apart for a weak trend
suggesting that far-IR luminous galaxies might have a somewhat
higher [CII]$E.W.$. We remind that the proposed calibration
in SFR of the [CII] line luminosity is valid for galaxies
spanning the range in far-IR luminosity
8.0 $\leq$ $\log L_{FIR}$ $\leq$ 10.50 L$_\odot$ and
0.2 $\leq$ [CII]$E.W.$ $\leq$ 3.0 $\mu$m.
It is thus difficult to predict whether the present calibration can be
applied to galaxies with $\log L_{FIR}$ $\geq$ 10.50 L$_\odot$.


\section {Discussion and conclusion}

The analysis carried out in the previous section has shown that the [CII] line
luminosity can be taken as a star formation indicator in normal,
late-type galaxies.
It is however clear that this result cannot be extended to far-IR ultraluminous
galaxies (ULIRG), where the ratio of the [CII] line over the continuum is a
factor of $\sim$ 10 lower than in normal late-type galaxies (Luhman et al.
1998).\\
The dispersion in the $L_{{\rm H}\alpha}$ vs. $L_{\rm [CII]}$ relation is a
factor of $\sim$ 3
(1 $\sigma$).
The uncertainty on the determination of the SFR of galaxies using relations
3 is however larger. As shown in Charlot \& Longhetti (2001), the
uncertainty on the determination of SFR from H$\alpha$ data using stellar
population synthesis models is already a factor of $\sim$ 3 when the
data are properly corrected for dust extinction and [NII] contamination. If we
take into account all the possible sources of error in the present
determination
of the H$\alpha$ luminosity, and the uncertainty that can be introduced
by the large dispersion observed in Fig. 1c, we can imagine that the
resulting uncertainty in the determination of the SFR from [CII] line
measurement is as high as a factor of $\sim$ 10 statistically. \\
While the calibration of the H$\alpha$ line luminosity in terms
of SFR is limited by the accuracy of photoionisation and stellar
population synthesis codes, and by the assumption that
the initial mass function (IMF) is universal, the uncertainty
introduced by the empirical calibration between $L_{\rm [CII]}$ and
$L_{{\rm H}\alpha}$ should be significantly reduced once larger samples of
galaxies with [CII] line data and H$\alpha$ imaging are available.
Part of this
  uncertainty may be due to the different mix of the contributions
  from different sources of the [CII] line emission within an individual
  galaxy and among galaxies of different Hubble types. We discuss this aspect
  in the Appendix.\\

\acknowledgements

We want to thank T. Onaka for providing us with unpublished LWS continuum data
for all galaxies listed in Negishi et al. (2001), and D. Bomans
for providing us with H$\alpha$ data of NGC 1569. A.B. thanks S. Madden
and A. Contursi for interesting discussions. We also thanks the referee,
Gordon Stacey, for his insight and suggestions.
This research has made use of
the NASA/IPAC Extragalactic Database (NED) which is operated by the Jet
Propulsion Laboratory, California Institute of Technology, under
contract with the National Aeronautics and Space Administration.

\noindent
{\bf Appendix}

In this Appendix we give a summary of our empirical knowledge of the
relations between the [CII] line emission and the dust emission in the
far-IR in
various parts of the interstellar medium, and we discuss how the
corresponding ratios compare with what is observed globally in galaxies.
This subject is a rather complex one and the results are liable to changes
when the data from ISO will be more fully analyzed. Thus this discussion is
only provisional.

There are several possible sources for the interstellar far-IR forbidden
lines, as discussed e.g. by Malhotra et al. (2001). Let us consider in
particular the principal cooling lines of the neutral gas, [OI] 63$\mu$m
and [CII] 158$\mu$m. While the [OI] line is nearly entirely emitted by dense
photodissociation regions (PDRs), the [CII] line is emitted not only by
these regions, but also by the diffuse neutral gas and by the ionized gas.
The far-IR continuum emission at wavelengths larger than about 100 $\mu$m is
due to big dust grains in thermal equilibrium, while at shorter wavelengths
there is a contribution from smaller grains heated out of equilibrium by
absorption of individual photons.

Let us now discuss briefly our knowledge of the [CII]/FIR ratios and, when
available, of the [CII]$E.W.$, for different components of the interstellar
medium.
The results are summarized in Table 3.

\medskip
\noindent
{\fontsize {11}{1}
{\sl {\sffamily A.1: Diffuse interstellar medium}}}
\medskip

\noindent
Here the published observations concern essentially the local diffuse
medium in the Solar neighbourhood, as observed at high galactic latitudes
in particular by COBE. From the correlation of the [CII] line intensity
$I([CII])$ with the column density of HI,  Bennett et al. (1994) find a [CII]
emission per H atom of the diffuse interstellar medium:

\begin{equation}
I_{\rm HI}([CII])/4\pi = (2.65 \pm 0.15)\, 10^{-26} \;  {\rm erg}\,{\rm
s}^{-1}\,{\rm H\, atom}^{-1}.
\end{equation}

\noindent
Together with a column density of HI of 3.3 10$^{20}$ csc $|b|$ atom cm$^{-2}$
(Lockman et al. 1986), this yields the following cosecant law for
the intensity of the [CII] line emitted by the neutral diffuse medium:

\begin{equation}
I_{\rm HI}([CII]) = (6.96 \pm 0.39) \, 10^{-7} \csc |b| \; {\rm erg}\,{\rm
cm}^{-2}\,{\rm
s}^{-1}\,{\rm sr}^{-1}.
\end{equation}

\noindent
From measurements of the
H$\alpha$ brightness at high galactic latitudes, Reynolds (1992) predicts
an emission by the diffuse warm ionized medium (WIM) of:

\begin{equation}
I_{\rm WIM}([CII]) = 3.6 \, 10^{-7} \csc |b| \; {\rm erg}\,{\rm
cm}^{-2}\,{\rm s}^{-1}\,{\rm
sr}^{-1},
\end{equation}

\noindent
in agreement with the upper limit
$I_{\rm WIM}([CII]) < 6\,10^{-7}$ erg cm$^{-2}$
s$^{-1}$ sr$^{-1}$ given by Bennett et al. (1994) at the galactic poles.
Reynolds (1992) also predicts for the intensity
$I([NII]_{205})$ of the [NII] 205$\mu$m line from the WIM:

\begin{equation}
I_{\rm WIM}([NII]_{205}) = 2.4 \, 10^{-8} \csc |b| \;
{\rm erg}\,{\rm cm}^{-2}\,{\rm s}^{-1}\,{\rm
sr}^{-1},
\end{equation}

\noindent
and a ratio between the intensities $I([NII]_{205})$ and $I([NII]_{122})$
of the [NII] 205$\mu$m and 122$\mu$m lines, respectively:

\begin{equation}
\frac{I_{\rm WIM}([NII]_{205})}{I_{\rm WIM}([NII]_{122})} = 1.4,
\end{equation}

\noindent
at the low-density limit. Bennett et al. (1994) observe
$I([NII]_{205}) = (4 \pm 1) \,
10^{-8} \csc |b| \; {\rm erg}\,{\rm cm}^{-2}\,{\rm s}^{-1}\,{\rm sr}^{-1}$ and
$I([NII]_{205})/I([NII]_{122}) \approx 0.7$. The comparison with the predicted
numbers,
given the uncertainties on the gaseous abundances of N and C and on the
observed intensities, leaves little doubt on the fact that the [NII] lines
at high galactic latitudes come from the WIM, and that the predicted value
of $I_{\rm WIM}([CII])$ is correct. The sum of $I_{\rm HI}([CII])$ and $I_{\rm
WIM}([CII])$ is however smaller than the total observed intensity of the
[CII] line, $I([CII]) = (1.43 \pm 0.12) \, 10^{-6} \csc |b| \; {\rm
erg}\,{\rm cm}^{-2}\,{\rm s}^{-1}\,{\rm sr}^{-1}$, but the difference might not
 be
significant given the uncertainties. To summarize, 2/3 of the [CII] line
emission
from the local diffuse medium comes from the neutral medium, and 1/3 comes from
the warm ionized medium. Lagache et al. (2000) show that the
FIR/submillimeter emissivity of dust associated with the two media is
approximately
the same, respectively $\tau/N({\rm HI}) = 8.3\,10^{-26}
(\lambda/250\mu{\rm m})^{-2}$ and $\tau/N({\rm H}^+) = 1.1\,10^{-25}
(\lambda/250\mu{\rm m})^{-2}$, with similar temperatures $T$ = 17.2 K. The
corresponding continuum brightness of the local diffuse medium at 158
$\mu$m is 4.94  csc$|b|$ MJy sr$^{-1}$, or 5.95 10$^{-7}$
csc$|b|$ erg cm$^{-2}$ s$^{-1}$ $\mu$m$^{-1}$ sr$^{-1}$.  This allows
one to calculate the equivalent width of the [CII] line from the local diffuse
medium, as 1.8 $\mu$m if we use for the [CII] line intensity the observed
contribution of the HI medium plus the estimated contribution of the WIM,
or 2.4 $\mu$m if we use the line intensity observed directly by Bennett et al.
(1994). The ratio of the [CII] line intensity to the far-IR intensity
(defined as
$I$(FIR) = 1.26 10$^{-14}$[2.58 $I$(60$\mu$m)/(1 Jy/sr) +
$I$(100$\mu$m)/(1 Jy/sr)]  W m$^{-2}$ sr$^{-1}$) is comprised
between 0.034 and 0.045; we used for this determination the spectral energy
distribution of the cirrus emission given by Fig. 3 of Boulanger et al. (2000).

We can simply estimate the variations of the [CII] line E.W. and of
$I$([CII])/$I$(FIR) with the the UV radiation field $G_0$ using the following
simplifying assumptions: $I$([CII]) is
proportional to $G_0$ (in units of the Habing field,
1.6 $\times$ 10$^{-3}$ erg~cm$^{-2}$s$^{-1}$, $G_0 \approx 1$ near the
Sun); the temperature of the big dust grains is $T_d = G_0^{1/6} \times
17.2$ K,
and their
emissivity is proportional to $\lambda^{-2}$ (see above); the emission of
the small
dust grains giving an excess radiation at 60 $\mu$m is proportional to $G_0$,
calibrated near the Sun using the cirrus spectral energy distribution of
Fig. 3 of
Boulanger et
al. (2000).  We find that $I$([CII])/$I$(FIR) {\it decreases} with increasing
$G_0$ in the range $1 < G_0 < 100$ (by a factor 2 at $G_0$ = 10 and a
factor 3 at $G_0$ = 100), while [CII]$E.W.$ {\it increases} by factors 1.8
and 5.4 respectively when $G_0$ increases from 1 to 10 and to 100.
The reason why the line to continuum ratio gets smaller with increasing $G_0$,
while the the equivalent width gets larger, is that the dust is getting
warmer,
shifting the far-IR continum to shorter wavelengths. More exact calculations
can be made using the models of Li \& Draine (2001) or of Dale et al. (2001),
but the results are not expected to be substantially different.\\
No good observational information exists on the
[CII]/FIR ratio in the diffuse medium of low-metallicity galaxies.
However a comparison of the [CII] map of the Large Magellanic Cloud of
Mochizuki et al. (1994) with the far-IR maps of Schwering (1989)
suggests $I$([CII])/$I$(FIR) $\simeq$ 0.010 $\pm$ 0.005 for the LMC.

\medskip
\noindent
{\fontsize {11}{1}
{\sl {\sffamily A.2: Weak PDRs}}}
\medskip

\noindent
Habart et al. (2001) give results from measurements with ISO on the PDR
associated with the molecular cloud L\,1721, illuminated by a B2 star, with
$G_0$
= 5 - 10.  The $I$([CII])/$I$(FIR) ratio is of the order of 0.012 and the
equivalent width of the [CII] line is about 6 $\mu$m. This is not very
different from what we just predicted above for $G_0$ = 10 (respectively
0.02 and 4 $\mu$m). This small discrepancy may be understood if the
density is lower than the
critical density for the [CII] line.

The southern molecular cloud N\,159S in the LMC is probably exposed to a
similar far-IR intensity. Israel et al. (1996) give $I$([CII])/$I$(FIR) $\geq$
0.025 for this cloud, somewhat higher than for L\,1721. This is the highest
value they quote for the LMC.

For the Orion giant molecular cloud at a one-degree scale, as described in
Stacey et al. (1993) (their one-degree scan), the $I$([CII])/$I$(FIR) ratio
is of the order of 0.003, with $G_0 \sim$ 25 (see their Sect. 4.2.3). This
might be considered as typical for giant molecular clouds from which OB stars
have recently formed.

\medskip
\noindent
{\fontsize {11}{1}
{\sl {\sffamily A.3: Strong PDRs}}}
\medskip

\noindent
Individual galactic strong and dense PDRs have values of $I$([CII])/$I$(FIR)
of the order of 5 10$^{-4}$ (Stacey et al. 1991b). These very low values are a
consequence of the saturation of the upper level of the 158 $\mu$m
transition at
high densities and radiation fields. An example is the 4\arcmin $\times$
4\arcmin ~``interface'' region near the Orion nebula, as defined and
discussed by
Stacey et al. (1993). Strong PDRs in the Magellanic Clouds have considerably
higher values of $I$([CII])/$I$(FIR), of the order of 0.015 (Israel et al.
1996).
As discussed by these authors, this exceptionally high value is associated
with
the low abundances of heavy elements in these galaxies. The effect of the low
metal abundances is both direct (less far-IR emission from dust) and indirect
(deeper penetration of UV light inside the PDRs) (see also Pak et al. 1998).
 However, due to the relatively poor linear resolution of the relevant
observations, some contribution for an associated giant molecular cloud heated
by a relatively weak UV radiation field might be included in the observing
beam,
raising the $I$([CII])/$I$(FIR) ratio as discussed previously for the Orion
region.

\medskip
\noindent
{\fontsize {11}{1}
{\sl {\sffamily A.4: Quiescent molecular clouds}}}
\medskip

\noindent
In a quiescent (i.e. with no or little star formation activity) molecular
cloud,
the [CII] line emission originates at the cloud
surface while the far-IR emission comes also from the cloud interior. In our
Galaxy, supposedly quiescent clouds
are observed by Stacey et al. (1991b) to have $I$([CII])/$I$(FIR) $\sim$
1.1 10$^{-3}$. They are not ordinary molecular clouds, however, but they are
expected to be bounded by relatively strong PDRs because they are in fact
immersed in a rather strong UV radiation field. Their higher
$I$([CII])/$I$(FIR)
ratio compared to that of strong PDRs simply suggests that the upper level
of the CII
transition is less saturated, probably due to a combination of lower
density and UV field. We have no information on the really quiescent molecular
clouds, but it is clear that in some circumstances molecular clouds heated
mainly by the visible and near-IR emission of relatively old stars can emit
in the far-IR without any emission in the [CII] line. This may be the case
for molecular clouds in the bulge of M\,31 (there is no direct
evidence for the star formation suggested by Schmidtobreick et al. (2000)
to be present inside some molecular clouds in this bulge). Similarly, regions
of large molecular complexes heated by the far-IR radiation of the surrounding
star-forming clouds (Falgarone \& Puget 1985) are expected to emit in the
far-IR but not in the [CII] line.

\medskip
\noindent
{\fontsize {11}{1}
{\sl {\sffamily A.5: HII regions}}}
\medskip

\noindent

Some fractions of the [CII] and far-IR global emissions may originate from
within HII regions. However, these contributions are negligible with
respect to those from other components of the ISM, on a galaxy scale. This
conclusion is suggested by the comparison
of the observed intensities of the [OIII], [NII], [CII] and [OI] lines  with
those predicted by photoionization models appropriate for HII regions, such as
those of Stasi\'nska (1990). It is then easy to see that the contribution
of HII
regions to the [CII], [NII] and [OI]  lines is negligible at the scale of a
galaxy. Similarly, the contribution of the dust inside the HII regions to the
overall far-IR flux is negligible.

\medskip
\noindent
{\fontsize {11}{1}
{\sl {\sffamily A.6: Comparison with galaxies}}}
\medskip

\noindent
On average, the values of $I$([CII])/$I$(FIR) and [CII]$E.W.$ observed for our
galaxies are of the order of 3 10$^{-3}$ and of 0.7 $\mu$m respectively, but
the dispersion is large.  These values are smaller than the corresponding
values
for the diffuse gas alone. For the irregular galaxies of Hunter et al. (2001),
$I$([CII])/$I$(FIR) = 2 - 6 10$^{-3}$. For the LMC as a whole,
$I$([CII])/$I$(FIR) = 0.01 (Israel et al. 1996). No conclusion can be reached
on the respective contributions of the different interstellar components for
the [CII] emission of the LMC, because the $I$([CII])/$I$(FIR) ratios are
similar within the (large) errors for PDRs and for the diffuse matter (but
see the discussion above on the strong PDRs in the Magellanic clouds). For
spiral galaxies, it is clear that the contributions of the PDRs to the
emission
of the [CII] line should be added to those of the diffuse neutral and ionized
media. However we have
seen that in these galaxies the strong PDRs exhibit values of
$I$([CII])/$I$(FIR) of about 5 10$^{-4}$. As a consequence, strong PDRs cannot
be the only contributors to the [CII] and far-IR emissions in normal spiral
galaxies: weak PDRs with $I$([CII])/$I$(FIR) of 0.003 to 0.015, and the
diffuse
medium with $I$([CII])/$I$(FIR) $\sim 0.04$, must also contribute to explain
the average ratio of 0.003. It is however possible to account for the
properties
of galaxies by a simple one-component analysis using
single values of the density and UV radiation field, as those of Hunter et al.
(2001), Malhotra et al. (2001) and Negishi et al. (2001). In such a model,
about
30,000 giant molecular clouds similar to the 1-degree Orion cloud of Stacey et
al. (1993), each one with a [CII] line luminosity of 1.5 10$^{-3}$
L$_{\odot}$,
would explain the both the luminosity of the [CII] line and the
$I$([CII])/$I$(FIR) ratio of our Galaxy. Such simple models might well be
justified for active star-forming galaxies, in which dense PDRs dominate by
far
the emission of the [CII] line. In more normal galaxies, however, all the
components we have described are likely to contribute to some extent and these
simple models have to be considered with caution. In some galaxies, the
contribution of low-density neutral gas to the [CII] line can reach as much
as
50\% of the total, as shown by Pierini et al. (2001). Unfortunately, it
does not
yet seem possible to disantangle the contributions of the various intestellar
components to the emission of the [CII] line and of the FIR continuum.

\begin{table}
\caption{Characteristics of the [CII] emission different galactic
and extragalactic environments.}
\label{Tab3}
\[
\begin{array}{lccc}
\hline
\noalign{\smallskip}
{\rm Component}            & {\rm G_0} & {\rm I([CII])/I(FIR)} & [CII]E.W.  \\
                           &           &                     &  {\rm \mu m} \\
\noalign{\smallskip}
\hline
\noalign{\smallskip}

{\rm Diffuse~gas}          &  1        &     0.04            &     2      \\
                           & 10        &     0.02            &     4      \\
{\rm Diffuse~gas~in~LMC}   & 10        &\simeq 0.01          &     -      \\
\hline
{\rm Weak~PDR}             & 3 - 10    &    0.012            &     6      \\
{\rm Weak~PDR~in~LMC}      & 10        &  > 0.02             &     -      \\
{\rm Orion~molec.~complex} & 25        &    0.03             &     -      \\
\hline
{\rm Strong~PDR}           &10^3-10^5  &   5~10^{-4}         &     -      \\
{\rm Strong~PDR~in~LMC}    &10^3-10^5  &     0.02            &     -      \\
\hline
{\rm Normal~spiral~galaxies}&  -        &   3~10^{-3}         &    0.7     \\
{\rm Irregular~galaxies}   &  -        &   2-6~10^{-3}       &     -      \\
{\rm LMC}                  &  -        &     0.01            &     -      \\
\noalign{\smallskip}
\hline
\end{array}
\]
\end{table}

A multi-component analysis is probably not feasible at present due to
the lack of analyzed data, but it might become possible in the near future
considering not only the emission in the fine-structure lines and the dust
infrared
continuum, but also the emission in the CO lines. This does not prevent the
use
of the
[CII] line as an indicator of star formation, but shows that for the moment
this indicator can only be calibrated empirically as we are doing here.

\end{document}